# Ascertaining hydrogen-abstraction reaction efficiencies of halogenated organic compounds on electron ionization mass spectrometry


**Caiming Tang[1],\*, Jianhua Tan[2], Yujuan Fan[1,3], Xianzhi Peng[1]**

[1] *State Key Laboratory of Organic Geochemistry, Guangzhou Institute of Geochemistry, Chinese Academy of Sciences, Guangzhou 510640, China*

[2] *Guangzhou Quality Supervision and Testing Institute, Guangzhou, 510110, China*

[3] *University of Chinese Academy of Sciences, Beijing 100049, China*

\*Corresponding Author.

Tel: +86-020-85291489; Fax: +86-020-85290009; E-mail: CaimingTang@gig.ac.cn (C. Tang).




## ABSTRACT


H-abstraction reactions occurring on electron ionization mass spectrometry (EI-MS) are a long-standing and crucial topic in MS research. Yet some critical relevant mechanisms are controversial and ambiguous, and information about the EI-induced H-abstraction reactions of halogenated organic compounds (HOCs) is completely in the dark. This study provides a systematic investigation of H-abstraction reactions of HOCs taking place on EI source using $^{13}C_6$-hexachlorobenzene ($^{13}C_6$-HCB) and $^{13}C_6$-hexabromobenzene ($^{13}C_6$-HBB) as exemplary compounds by gas chromatography high resolution mass spectrometry (GC-HRMS). The H-abstraction efficiencies were evaluated with the MS signal intensity ratios of ions with H-abstraction relative to the corresponding original ions (without H-abstraction). Ion source temperatures, EI energies and numbers of heavy isotope atoms ($^{37}Cl$ or $^{81}Br$) of isotopologues were investigated in terms of their effects on the H-abstraction efficiencies. The H-abstraction efficiencies of individual isotopologues generally decreased from the first to the last isotopologues of respective ions, and those of individual ions were different from each other, with the highest values of 0.017 and 0.444 for $^{13}C_6$-HCB and $^{13}C_6$-HBB, respectively. The overall H-abstraction efficiencies involving all measured ions of $^{13}C_6$-HCB and $^{13}C_6$-HBB were 0.004 and 0.128, respectively. With increasing ion source temperatures, the H-abstraction efficiencies firstly increased to a summit and then began to linearly decrease. EI energies and emission currents could impact the H-abstraction efficiencies, but showed no certain tendency. The H-abstraction reactions were inferred to belong to ion-molecule reactions, and the siloxanes bleeding from the GC column might be the hydrogen source. Some strategies were proposed




for eliminating or alleviating the interference triggered by the H-abstraction reactions on EI-MS in identification of halogenated organic pollutants (HOPs). Our findings provide a better understanding for the EI-induced H-abstraction reactions of HOCs, and may benefit identification of HOPs in environmental analysis, especially for novel HOPs.

## Keywords:





# 1 INTRODUCTION

Hydrogen abstraction (H-abstraction) reactions (or intermolecular hydrogen transfer reactions), which may harbor important significance for mass spectrometry (MS) analysis, can commonly take place on various MS ionization sources, such as electron ionization (EI),[1,2] chemical ionization (CI),[3] electrospray ionization (ESI), [4-8] atmospheric pressure chemical ionization (APCI),[9,10] matrix-assisted laser desorption ionization,[11] atmospheric pressure photoionization (APPI) [12-14] and low-pressure photoionization (LPPI) sources. [15-18] The mechanisms of H-abstraction reactions on CI, ESI, APCI and photoionization (PI) sources have been well studied. However, the real mechanisms of H-abstraction reactions on EI-MS remain far from completely explicit, although the H-abstraction phenomenon has been observed and investigated for a very long time (2-5 decades).[1,2,19-24]

It was reported that the generation of molecular ions with H-abstraction on EI source, such as $[M+H]^+$, $[M+2H]^+$ and $[M+3H]^+$, might be related to the structures of analytes and the ion source temperatures.[23,24] Some studies concluded that the transferred hydrogen atoms in H-abstraction reactions on EI source were derived from the $H_2O$ in ion source and/or samples.[1,2,21,25,26] However, a study revealed that the hydrogen atoms did not stem from water residual in the ion source, neither $H_2$ residual.[23] The author inferred that the hydrogen source was related with thermal initiation and EI ionization.[23] Most of the previous studies indicated that the H-abstraction reactions on EI source belonged to ion-molecule reactions, [1,2,19,22-27] which can be expressed as:



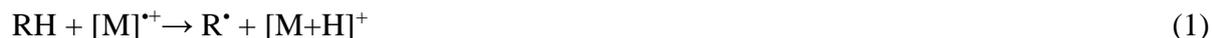

$$RH + [M]^{\bullet+} \rightarrow R^{\bullet} + [M+H]^{+} \tag{1}$$

This reaction equation is similar to that of some H-abstraction reactions occurring on PI source.[9,28-31] The initial ionization process of PI is:[9,29-30]

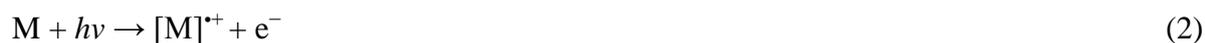

$$M + h\nu \rightarrow [M]^{\bullet+} + e^{-} \tag{2}$$

and that of EI is:

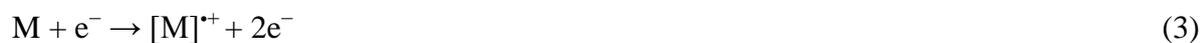

$$M + e^{-} \rightarrow [M]^{\bullet+} + 2e^{-} \tag{3}$$

It can be seen from Eqs (2) and (3) that the initial ionization processes of PI and EI are to some extent similar. As a result, the revealed mechanisms behind the H-abstraction reactions occurring on PI source may provide illuminating insights into the H-abstraction fundamentals on EI source. On PI source, the hydrogen source for H-abstraction reactions is protic solvents used for injection, such as toluene,[29] water,[32] methanol,[33] ethanol and hexane.[28] Whereas aprotic solvents such as tetrachloromethane and aprotic gases including $N_2$, $O_2$ and $CO_2$ could compete with analytes to abstract hydrogen atoms, thus impair the MS signal intensities of $[M+H]^{+}$.[28] In addition to solvents, the temperature, ionizer voltage and reaction length of PI source also play critical roles in the H-abstraction reactions,[28,29,34] showing positive correlations with the H-abstraction efficiencies. Furthermore, the proton affinities of analytes and the reaction enthalpy changes ($\Delta H$) are also important factors affecting the H-abstraction reactions on PI source, in positive correlation with the H-abstraction efficiencies.[28,34]



Up to now, reported studies concerning H-abstraction reactions on EI source mainly focused on quinones.[20,21] While halogenated organic compounds (HOCs), e.g., organochlorines and organobromines, may also abstract hydrogen atoms on EI source, in view of their lone electron pairs of halogen atoms (e.g., Cl and Br). HOCs are a large category of chemicals produced by both natural and anthropogenic activities,[35-38] and have been contributing enormous influences to human beings and the environment, in either positive or negative ways.[39-43] Many HOCs are persistent, bioaccumulative and toxic pollutants, such as polychlorinated dioxins/furans,[44-46] dishlorodiphenyltrichloroethanes (DDTs),[47-49] polychlorinated biphenyls,[45,50] polybrominated diphenyl ethers,[48,49] polybrominated biphenyls,[51] chlorinated paraffins,[52] perfluorinated compounds[53] and etc., which have been posing serious pollution to the environment. In addition to known halogenated organic pollutants (HOPs), numerous unknown, novel and emerging HOPs have been continuously screened and identified in various matrices.[54-58] Most HOPs can be analyzed by gas chromatography coupled with EI-MS (GC-EI-MS), which is a main stream approach for analysis of HOPs.[59] Recently, we successfully developed a quasi-targeted analysis method for identification of novel HOPs in environmental matrices using GC-EI-HRMS.[60] Additionally, some non-hyphenated MS techniques such as EI-FTICR-HRMS have been applied to identification of HOPs.[61] Nevertheless, anticipatable H-abstraction reactions on EI-MS may introduce plausible interference peaks and thereby could compromise the identification of HOPs, especially non-targeted or quasi-targeted analysis of unknown and novel pollutants. Accordingly, the H-abstraction reactions of HOCs on EI-MS require systemic investigation to facilitate the screening and identification of these compounds.



In this study, we conducted an in-depth study concerning the H-abstraction reactions of HOCs occurring on EI-MS using GC-double focusing magnetic-sector HRMS (GC-DFS-HRMS), with two stable-isotope labelled standards, i.e., $^{13}C_6$-hexachlorobenzene ($^{13}C_6$-HCB) and $^{13}C_6$-hexabromobenzene ($^{13}C_6$-HBB), used as model compounds. The H-abstraction efficiencies were calculated as the ratios of MS signal intensities between ions with H-abstraction and the corresponding original ions (without H-abstraction), and the effects of ion source temperatures, EI energies as well as numbers of heavy isotope atoms ($^{37}Cl$ or $^{81}Br$) of isotopologues on the H-abstraction efficiencies were investigated. This study reveals some details of H-abstraction reactions of HOCs on EI-MS, provides new insights into the potential mechanisms of the EI-induced H-abstraction reactions, and may be conducive to non-targeted and quasi-targeted qualitative analyses of HOPs, particularly novel HOPs.



## 2 EXPERIMENTAL

### 2.1 Chemicals and materials

Stock solutions of stable-isotope labelled standards $^{13}C_6$-HCB (100 µg/mL in nonane) and $^{13}C_6$-HBB (100 µg/mL in toluene) were purchased from Cambridge Isotope Laboratories Inc. (Andover, MA, USA). Structures and CAS numbers of the standards are documented in Table S1. Chromatographic-grade isooctane was purchased from CNW Technologies GmbH (Düsseldorf, Germany). Acetone and n-hexane used for washing the auto-sampler injection syringe were bought from Merck Crop. (Darmstadt, Germany). Perfluorotributylamine used as a reference standard for HRMS calibration was purchased from Sigma-Aldrich LLC. (St. Louis, MO, USA). The purchased stock standards of $^{13}C_6$-HCB and $^{13}C_6$-HBB were further diluted with isooctane to prepare working solutions at the concentrations of 5 µg/mL and 1 µg/mL, respectively. All the standard solutions were kept in a fridge at −20 ºC prior to use.

### 2.2 GC-HRMS analysis

The GC-HRMS system consisted of dual gas chromatographs (Trace-GC-Ultra) coupled to a double focusing magnetic-sector HRMS (BE geometry) and an extra TriPlus auto-sampler (GC-DFS-HRMS, Thermo-Fisher Scientific, Bremen, Germany). The prepared working solutions were directly analyzed by the GC-HRMS. The chromatographic separation was conducted with a DB-5MS capillary column (60 m × 0.25 mm, 0.25 µm film thickness, J&W Scientific, Folsom, USA).



The GC temperature-ramping program is documented as follows: held at 120 ℃ for 2 min; ramped to 220 ℃ at 20 ℃/min, held for 16 min; then ramped to 235 ℃ at 5 ℃/min, held for 7 min; and then ramped to 260 ℃ at 5 ℃/min; and finally ramped to 310 ℃ at 40 ℃/min and held for 0.75 min. The total time of a GC run was 40 min. The solvent delay time was 6.5 min. Ultra-high purity helium was used as the carrier gas and maintained at a constant flow rate of 1.0 mL/min. The GC injection port and transfer line were kept at 260 ℃ and 280 ℃, respectively. Splitless injection mode was applied and the injection volume was 1 μL.

The working settings and parameters of the DFS-HRMS are provided as the following: EI source operated in positive mode was applied; EI energies were set at 35, 45, 55 and 65 eV; ion source temperatures were set at 160, 190, 220, 250, 280 and 300 ℃; filament emission currents ranged from 0.463 to 0.989 mA with the corresponding EI energies from 35 to 65 eV; data acquisition was performed with multiple ion detection (MID) mode; dwell time was 20 ms for each isotopologue; scan time segments were set at 8-20 min (for $^{13}C_6$-HCB) and 20-36 min (for $^{13}C_6$-HBB); mass resolution (5% peak definition) was ≥ 10000 and the MS detection accuracy was set at ±1 mu. If not specified, all measurements were performed with the EI energy of 45 eV and the ion source temperature at 250 ℃. The HRMS real-time calibration was carried out with perfluorotributylamine during MID operation.

Chemical structures of the investigated compounds along with their dechlorination/debromination radicals were depicted with ChemDraw (Ultra 7.0, Cambridgesoft), and the exact masses of the chlorine/bromine isotopologues were calculated



with mass accuracy of 0.00001 u. Since the carbon of $^{13}C_6$-HCB and $^{13}C_6$-HBB comprises only

$^{13}C$ isotope, thereby only chlorine/bromine isotopologues were studied. Hydrogen ($^1H$) was the

only involved hydrogen isotope, since others (deuterium and tritium) have extremely lower

abundances in the nature relative to hydrogen. For a molecule or radical containing $n$ Cl/Br

atoms, all the isotopologues ($n+1$) were chosen. Exact mass-to-charge ratios ($m/z$) of ions were

calculated by subtracting the exact mass of one electron from that of individual isotopologues.

The $m/z$ values of ions with H-abstraction were obtained by adding the exact mass of a hydrogen

atom to that of each original ion, and only the H-abstraction reactions transferring merely one

hydrogen atom were taken into consideration. These $m/z$ values were imported into the MID

module for detection. The detailed information including chemical formulas, exact masses and

exact $m/z$ values of the isotopologues of the investigated compounds is listed in Table S2, and

the representative chromatograms along with high-resolution mass spectra of the compounds

are shown in Figure 1.

## 2.3 Data processing

The H-abstraction efficiency of an isotopologue was expressed as the MS signal intensity ratio

of the isotopologue with H-abstraction relative to the original isotopologue ($R_i$), and calculated

by

$$R_i = I'_i / I_i \qquad (4)$$

where $i$ is the number of $^{37}Cl$ or $^{81}Br$ atom(s) in the isotopologue; $I'_i$ is the MS signal intensity



of the isotopologue $i$ with H-abstraction; $I_i$ is the MS signal intensity of the original

isotopologue $i$. The whole H-abstraction efficiency of an ion ($R_n$) was calculated with

$$R_n = \sum_{i=0}^{n} I'_i \bigg/ \sum_{i=0}^{n} I_i \qquad (5)$$

where $n$ is the number of Cl/Br atoms of the ion. In addition, the overall H-abstraction efficiency

involving all the detected ions of a compound ($R_{all}$) was calculated with

$$R_{all} = \sum_{n=1}^{6} \sum_{i=0}^{n} I'_i \bigg/ \sum_{n=1}^{6} \sum_{i=0}^{n} I_i \qquad (6)$$

Prior to exporting MS signal intensities, background subtraction was carried out by deducting

baseline signal intensities close to both ends of individual chromatographic peaks. If not

specified, the number of injection replicates was six, and the experimental data derived from

these replicated measurements were applied to calculation of average H-abstraction efficiencies

and standard deviations ($1\sigma$). In order to describe this work more concisely and clearly, we use

IST-n to refer to an isotopologue with $n-1$ $^{37}$Cl or $^{81}$Br atom(s), apply P-Cl$_n$ to denote a

dechlorination product ion having n Cl atom(s), and utilize P-Br$_n$ to represent a debromination

product ion containing n Br atom(s). Because the IST-1 of P-Cl$_1$ with H-abstraction of $^{13}$C$_6$-

HCB ($[^{13}$C$_6$H$^{35}$Cl]$^+$, $m/z$: 113.99626) was seriously interfered by the $[C_2NF_4]^+$ ion ($m/z$:

113.99614) of perfluorotributylamine, the H-abstraction efficiency of the IST-1 was not

calculated.

## 2.4 Method performances



As documented in Tables S-3, S-4 and S-5, the precisions (standard deviations) of the H-abstraction efficiencies of individual isotopologues, the whole H-abstraction efficiencies of individual ions, and the overall H-abstraction efficiencies of individual compounds are within the ranges of 0.00002-0.01608, 0.00011-0.00871 and 0.00015-0.00511, respectively. These results demonstrate that the analysis method could meet requirements for investigating the H-abstraction efficiencies of $^{13}C_6$-HCB and $^{13}C_6$-HBB on EI-MS.

## 2.5 Statistical analysis

Statistical analysis was conducted with SPSS Statistics 19.0 (IBM Inc., Armonk, USA) and Origin 9 (OriginLab Corp., Northampton, USA). Paired-samples T test and independent-samples T test (performed with SPSS) were applied to determining p-values (2-tailed) with alpha of 0.01 as the threshold value for significance. The differences of the H-abstraction efficiencies and MS signal intensities among isotopologues were examined with the paired-samples T test, and those of the whole H-abstraction efficiencies among ions and the overall H-abstraction efficiencies between the two compounds were evaluated with the independent-samples T test. If a p-value is less than 0.01, the null hypothesis (e.g., no difference between two groups of H-abstraction efficiencies) is rejected, indicating an indeed existent significant difference. Linear and non-linear regressions (conducted with Origin) were employed to reveal the relationships between experimental data and parameters/factors of interest.



# 3 RESULTS AND DISCUSSION

## 3.1 Measured H-abstraction efficiencies

Isotope-labeled $^{13}C_6$-HCB and $^{13}C_6$-HBB instead of native HCB and HBB were chosen to perform this study, owing to that $^{13}C_6$-HCB and $^{13}C_6$-HBB contain only $^{13}C$ and Cl/Br atoms and thereby the interference caused by carbon isotopes is nonexistent. Based on the detected MS signal intensities of the original ions and those with H-abstraction, we calculated three hierarchical H-abstraction efficiencies, i.e., H-abstraction efficiencies of individual isotopologues, whole H-abstraction efficiencies of individual ions, and overall H-abstraction efficiencies of individual compounds. Thus, H-abstraction efficiencies of the investigated compounds were evaluated at three levels, namely, isotopologues, ions and compounds.

### 3.1.1 H-abstraction efficiencies of individual isotopologues

As shown in Figure 2 and Table S3, the first six chlorine isotopologues of the molecular ion of $^{13}C_6$-HCB, i.e., IST-1 to IST-6, were possessed of the H-abstraction efficiencies within the range from 0.00010±0.00009 to 0.0026±0.0003, exhibiting a gradual decline from the first to the last isotopologues (Figure 2a). For the first five chlorine isotopologues of P-Cl$_5$ (IST-1 to IST-5), the H-abstraction efficiencies were from 0.016±0.001 to 0.0184±0.0003, showing a generally decreasing tendency from IST-1 to IST-5 (Figure 2b). The H-abstraction efficiencies of individual chlorine isotopologues of other four dechlorination product ions (P-Cl$_1$ to P-Cl$_4$) ranged from 0.0020±0.0001 to 0.010±0.001 (Figure 2c-2f), presenting no explicit variation



trend along with the isotopologue sequences with statistical significance (Figure 2c-2e). The H-abstraction efficiencies of IST-1 of the molecular ion and the P-Cl$_5$ were significantly higher than those of other isotopologues of the respective ions (p ≤ 0.001). The H-abstraction efficiency difference between any two neighboring chlorine isotopologues of the molecular ion was statistically significant (p ≤ 0.008), and that between any two neighboring isotopologues of the first three isotopologues of the P-Cl$_5$ was statistically significant as well (p ≤ 0.001).

The H-abstraction efficiencies of the bromine isotopologues of the molecular ion of $^{13}C_6$-HBB decreased continuously from IST-1 (0.222±0.009) to IST-7 (0.0005±0.0005) (Figure 2g), and those of any neighboring isotopologues showed statistically significant difference (p ≤ 0.00001). Likewise, the product ions P-Br$_5$, P-Br$_4$, P-Br$_3$ and P-Br$_2$ of $^{13}C_6$-HBB presented gradually descending H-abstraction efficiencies from the first isotopologues to the last ones, with the H-abstraction efficiency ranges of 0.424±0.016 to 0.569±0.006, 0.111±0.007 to 0.23±0.01, 0.173±0.003 to 0.195±0.003, and 0.041±0.001 to 0.059±0.005, respectively (Figure 2h-2k). For these product ions, the H-abstraction efficiencies of the bromine isotopologues of each ion (except the IST-4, IST-5 and IST-6 of P-Br$_5$) can be confidently differentiated from each other (p ≤ 0.002). In addition, there was a slight decline for the H-abstraction efficiencies from IST-1 to IST-2 of the P-Br$_1$ (p ≤ 0.004) (Figure 2l).

### 3.1.2 Whole H-abstraction efficiencies of individual ions

As shown in Figure 3a, the magnitudes of the measured whole H-abstraction efficiencies of individual ions of $^{13}C_6$-HCB follow the order as: P-Cl$_5$ > P-Cl$_1$ > P-Cl$_4$ > P-Cl$_3$ > P-Cl$_2$ >



molecular ion, with the range from 0.0011±0.0001 to 0.0172±0.0002 (Table S4). These H-abstraction efficiencies are statistically distinguishable (p ≤ 0.00004). The scales of the whole H-abstraction efficiencies of individual ions of $^{13}C_6$-HBB showed the following order: P-Br$_5$ > P-Br$_3$ > P-Br$_1$ > P-Br$_4$ > P-Br$_2$ > molecular ion (Figure 3b), with the range from 0.037±0.001 to 0.444±0.009 (Table S4). These H-abstraction efficiencies can also be confidently differentiated (p ≤ 0.0003). The ion P-Br$_5$ presented the extremely high H-abstraction efficiency (0.444±0.009), suggesting that around 30.7% of the amount of the initial ion was subjected to H-abstraction reaction. It is noteworthy that the patterns of the whole H-abstraction efficiencies of ions of the two compounds are somewhat similar, probably implying similar mechanisms of H-abstraction reactions for the detected ions between the two compounds on EI-MS.

### 3.1.3 Overall H-abstraction efficiencies of individual compounds

As Figure 3c and Table S5 show, the overall H-abstraction efficiency of $^{13}C_6$-HBB (0.128±0.005) was over one order of magnitude higher than that of $^{13}C_6$-HCB (0.0044±0.0001), with the p-value < 0.00001. This result indicates that $^{13}C_6$-HBB was in possession of significantly higher H-abstraction capability in comparison with $^{13}C_6$-HCB on EI-MS.

### 3.2 Distribution patterns of measured MS signal intensities of molecular isotopologues

As shown in in Figure 4 and Table S6, it is very surprising that the distribution patterns of the measured MS signal intensities of isotopologues between the molecular ion ([M]$^{•+}$) and the molecular ion with H-abstraction ([M+H]$^+$) of each compound are apparently different. For the



[M]$^{\bullet+}$ of $^{13}C_6$-HCB, the measured MS signal intensities were in the following order: IST-2 > IST-3 > IST-1 > IST-4 > IST-5 > IST-6 > IST-7 (Figure 4a), which is identical to the theoretical relative abundance order. Whereas the detected isotopologue abundances of the [M+H]$^+$ of $^{13}C_6$-HCB presented the order as: IST-1 > IST-2 > IST-3 > IST-4 > IST-5 > IST-6 > IST-7 (Figure 4b). The detected abundance of IST-1 of the [M+H]$^+$ was significantly higher than those of IST-2 and IST-3 ($p \leq 0.00005$), which contradicts the scenario of the [M]$^{\bullet+}$ and thereby the order of theoretical relative abundances of isotopologues.

In addition to $^{13}C_6$-HCB, $^{13}C_6$-HBB also showed different measured isotopologue abundance orders between the [M]$^{\bullet+}$ and [M+H]$^+$. For the [M]$^{\bullet+}$, the order of the measured isotopologue abundances was: IST-4 > IST-3 > IST-5 > IST-2 > IST-6 > IST-1 > IST-7 (Figure 4c), which is well consistent with normal distribution and the theoretical relative abundance order. However, the measured isotopologue abundances of the [M+H]$^+$ exhibited the following order: IST-3 > IST-2 > IST-4 > IST-1 > IST-5 > IST-6 > IST-7 (Figure 4d), which is completely different from that of the [M]$^{\bullet+}$ and accordingly the theoretical relative abundance order. These findings also point to the significant differences of H-abstraction efficiencies among different chlorine/bromine isotopologues on EI-MS.

### 3.3 Parameters and factors affecting H-abstraction efficiency variation

To investigate whether some parameters can influence the H-abstraction efficiencies on EI-MS, we measured the H-abstraction efficiencies of the molecular ion of $^{13}C_6$-HCB with varied ion source temperatures and EI energies. Some previous studies have proved that ion source



temperatures could dramatically impact the extents of H-abstraction reactions on EI-MS, positively correlating with the reaction extents.[23,24] In the present study, however, we observed a different variation mode of H-abstraction efficiencies with ion source temperatures in contrast with those found in the previous studies. As shown in Figure 5a and Table S7, the measured H-abstraction efficiencies firstly increased from 0.0071±0.0008 to 0.0088±0.0003 with the ion source temperatures changing from 160 °C to 190 °C, and then almost linearly decreased to 0.0008±0.0001 with the temperatures varying from 190 °C to 300 °C. As Ahmed et al. have reported that the relationship between the H-abstraction reaction extents and the APPI source temperatures was an Arrhenius-type curve,[29] we accordingly plotted the measured H-abstraction efficiencies versus the reciprocal Kelvin temperatures. However, we did not get an Arrhenius-type temperature-dependent curve describing the relationship between the measured H-abstraction efficiencies and the ion source temperatures. Instead, the H-abstraction efficiencies first linearly increased from the lowest value to a summit with the 1/temperature values varying from 0.00174 $K^{-1}$ to 0.00216 $K^{-1}$ and then began to decrease as the 1/temperature ascended to 0.00231 $K^{-1}$ (Figure 5b).

As indicated in Figure 5c and 5d, the H-abstraction efficiencies can be significantly influenced by EI energies and emission currents of ion source filament. At the EI energy of 45 eV, the H-abstraction efficiency reached the highest (0.0044±0.0007), and significantly higher than the rest (p ≤ 0.003). While the H-abstraction efficiency at the EI energy of 35 eV was the lowest (0.0007±0.0002), and significantly lower than those at other EI energies (p ≤ 0.01). The H-abstraction efficiencies at EI energies of 55 eV and 65 eV were relatively comparable



(0.0014±0.0002 and 0.00168±0.00001), and cannot be differentiated with statistical significance (p = 0.2). Since the emission currents generally increased with the increase of EI energies, the order of the H-abstraction efficiencies observed at different emission currents were the same to that of the H-abstraction efficiencies at the corresponding EI energies. However, the pattern of H-abstraction efficiencies plotted versus EI energies is apparently different from that of H-abstraction efficiencies versus emission currents (Figure 5c and 5d). This observation might be ascribed to that the emission currents varied with the variation of EI energies more drastically at lower energy regions than at higher energy regions.

As Figure 5e shows, when the MS signal intensity variation was caused by the variation of ion source temperatures, a strong positive linear correlation between H-abstraction efficiencies and MS signal intensities could be observed ($R^2$=0.969). Nevertheless, when the MS signal intensity variation was triggered by the variation of EI energies, only a weak positive linear correlation between H-abstraction efficiencies and MS signal intensities was found ($R^2$=0.274, Figure 5f). This result demonstrates that the MS signal intensity may be a parameter affecting the H-abstraction efficiencies of organochlorines on EI-MS.

As illustrated in Figure 6, the H-abstraction efficiencies of isotopologues generally decreased with the increase of heavy isotope atoms ($^{37}$Cl or $^{81}$Br) of the isotopologues. The relationships between the H-abstraction efficiencies and the numbers of $^{37}$Cl or $^{81}$Br atom(s) were well fitted with exponential curves for the molecular ions and the P-Cl$_5$/P-Br$_5$ of $^{13}$C$_6$-HCB and $^{13}$C$_6$-HBB, with the $R^2$ values ranging from 0.975 to 0.998. In addition to the ions involved in Figure 6, the



ions P-Br$_4$, P-Br$_3$ and P-Br$_2$ also exhibited exponentially descending H-abstraction efficiencies of bromine isotopologues with the increase of $^{81}$Br atoms of the isotopologues, with R$^2 \geq 0.982$ (Figure S1). These findings suggest that chlorine/bromine isotopic effects may play a crucial role in the variation of H-abstraction efficiencies among chlorine/bromine isotopologues of HOCs on EI-MS.

### 3.4 Tentative mechanistic interpretation

### 3.4.1 Ion-molecule reactions and hydrogen source

Previous studies have concluded that the H-abstraction reactions on EI source are ion-molecule reactions,[1,2,19,22-27] which can be expressed as Eq (1). Prior to the H-abstraction reactions, the molecules of analytes are ionized as Eq (3) shows, giving rise to molecular and fragmental ions, which are electron-deficient radical ions. In light of the similar ionization processes of PI and EI as well as the reported outcomes,[28,29] we infer the most predominant ion-molecule reactions are that as shown by Eq (1), instead of others such as:

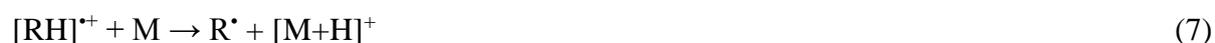

$$[RH]^{\bullet+} + M \rightarrow R^{\bullet} + [M+H]^{+} \tag{7}$$

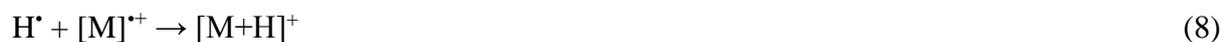

$$H^{\bullet} + [M]^{\bullet+} \rightarrow [M+H]^{+} \tag{8}$$

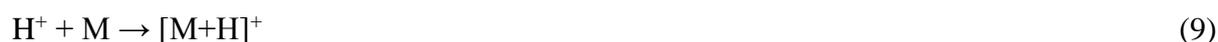

$$H^{+} + M \rightarrow [M+H]^{+} \tag{9}$$

On the other hand, Eq (7) is energetically unfavorable compared to Eq (1), while Eqs (8) and (9) are low probability events relative to Eq (1) due to the production of H$^{\bullet}$ and H$^{+}$ is not



energetically favorable. In addition, we found that not only the molecular ions but also the

dehalogenation fragmental ions participated in the H-abstraction reactions on EI source, and

the fragmental ions exhibited evidently higher H-abstraction efficiencies than the molecular

ions (Figure 3). This result manifests that the "molecules" in the ion-molecule reactions were

not the molecules of analytes, and the "ions" in the reactions were the molecular and fragmental

ions of the analytes, which is consistent with Eq (1). Then, the subsequent question is what the

real "molecules" in the ion-molecule reactions are or where the hydrogen atoms stem from.

We inferred that the reaction molecules might be siloxanes which ran off from the packing

materials (95% polydimethylsiloxane and 5% phenyl group) of the GC column and/or

emigrated from the silicone inlet septum of the GC.[62] Therefore, the H-abstraction reactions

can be described as:

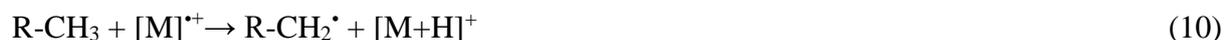

$$R\text{-}CH_3 + [M]^{\cdot+} \rightarrow R\text{-}CH_2^{\cdot} + [M+H]^+ \tag{10}$$

where R represents a replaceable siloxane group. It has been reported that the methyl of toluene

is preferred to contribute hydrogen atoms for H-abstraction reactions on PI source in contrast

with the phenyl group.[13] Accordingly, it can be anticipated that the methyl of siloxanes is an

excellent hydrogen donor for the H-abstraction reactions on EI source. In conclusion, the

siloxanes bleeding from the GC column and/or injection port pad might take participation in

the H-abstraction reactions on EI source and could be the hydrogen source.



Another evidence for the above conclusion is the difference of overall H-abstraction efficiencies between $^{13}C_6$-HCB and $^{13}C_6$-HBB, i.e., $^{13}C_6$-HCB presented significantly lower H-abstraction efficiency compared to $^{13}C_6$-HBB (Figure 3c). The electronegativity of chlorine is higher than that of bromine, thus the radical ions of $^{13}C_6$-HCB are more competent to abstract hydrogen atoms from hydrogen donors. However, in the present study, we found the overall H-abstraction efficiency of $^{13}C_6$-HCB is significantly lower than that of $^{13}C_6$-HBB. This implies that the densities of gaseous molecules of the hydrogen donors during the H-abstraction reactions were different for $^{13}C_6$-HCB and $^{13}C_6$-HBB. If not in this case, the H-abstraction efficiency of $^{13}C_6$-HCB ought to be higher than that of $^{13}C_6$-HBB. It can be inferred that the densities of gaseous molecules of the hydrogen donors for H-abstraction reactions of $^{13}C_6$-HCB were lower than those for H-abstraction reactions of $^{13}C_6$-HBB. This inference is in line with the reality that $^{13}C_6$-HCB eluted from the GC column at a relatively lower temperature than $^{13}C_6$-HBB and the higher oven temperature could cause more amount of siloxanes bleeding from the column.

### 3.4.2 Temperature-dependent reactions

In previous studies, the H-abstraction efficiencies on EI source were found to exponentially correlate with the ion source temperatures,[23,24] which was also observed in another study concerning the H-abstraction reactions on PI source.[29] Nevertheless, in our study, the H-abstraction efficiencies rose first and then fell with the ascending temperatures of ion source (Figure 5a). From 160 ℃ to 190 ℃, the H-abstraction efficiencies rose to a summit, which is consistent with the findings of the previous reports.[23,24,29] Whereas the H-abstraction



efficiencies gradually decreased to the lowest from 190 ºC to 300 ºC, which contradicts the outcomes of the previous studies.[23,24,29] We deduce that the decline of H-abstraction efficiencies with the increase of ion source temperature might be attributable to the thermostability of $[M+H]^+$. The pyrolysis rate of $[M+H]^+$ rises as the temperature increases. On the other hand, the rate of H-abstraction reaction generating $[M+H]^+$ may also increase along with the increase of ion source temperature. When the pyrolysis rate and generation rate of $[M+H]^+$ are equivalent, the H-abstraction efficiency reaches the highest.

### 3.4.3 Chlorine and bromine isotopic effects in H-abstraction reactions

The ion-molecule H-abstraction reactions of $^{13}C_6$-HCB and $^{13}C_6$-HBB on EI source may be analogous to H-abstraction reactions of chlorine and bromine radicals. We hypothesize a diatomic hydride HA (A is an imaginary element), and the H-abstraction reaction between HA and Cl$^•$ in gas phase is:

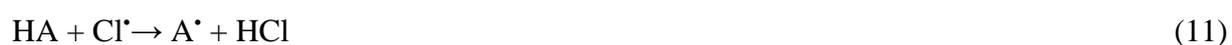

$$HA + Cl^• \rightarrow A^• + HCl \tag{11}$$

The transition-state complex is A--H--Cl. While taking the chlorine isotopes into account, the transition-state complexes become A--H--$^{35}$Cl and A--H--$^{37}$Cl. Due to zero point energies, the bond H--$^{35}$Cl is slightly weaker than the bond H--$^{37}$Cl. As a consequence, the energy barrier of A--H--$^{35}$Cl is slightly lower than that of A--H--$^{37}$Cl. Then, the formation of A--H--$^{35}$Cl is a little faster than that of A--H--$^{37}$Cl in the reaction as indicated in Eq (11), thereby generating more amount of H$^{35}$Cl in comparison with H$^{37}$Cl. Extending this inference to the ion-molecule H-



abstraction reactions of $^{13}C_6$-HCB on EI source, we conclude that the production of $[M+H]^+$ or $[P+H]^+$ of the compound was slightly higher for the lighter chlorine isotopologues than for the heavier ones. Analogously, it can be deduced that the yields of ions with H-abstraction on EI source of $^{13}C_6$-HBB were a little higher for the lighter bromine isotopologues than for the heavier ones. These theoretical inferences are in agreement with the experimental outcomes of this study (Figure 6 and Figure S1).

It is noteworthy that the correlations between the H-abstraction efficiencies and the $^{37}Cl$ or $^{81}Br$ atom numbers of chlorine/bromine isotopologues exhibit well-fitted Arrhenius-type curves (Figure 6 and Figure S1). This observation implies that the H-abstraction reactions on EI source were activation-energy-dependent (kinetically controlled) and the chlorine and bromine isotope effects played a part in the reactions.

### 3.5 Implications for EI-MS analysis of HOPs

Nowadays, non-targeted and quasi-targeted analyses as well as suspect screening have become feasible identification methods for known and unknown HOPs.[60] The characteristic isotopologue distributions are always applied to screening HOPs of interest.[60,63-65] The comparison between the theoretical isotopologue distributions and the detected mass spectra of isotopologues can facilitate the identification of HOPs, particularly novel HOPs, thus promoting the reliability of identification results. Even so, it remains unclear whether the H-abstraction reactions on EI source can affect the identification reliability of HOPs when isotopologue distribution comparison is applied. For instance, the ion P-Cl$_5$ with H-abstraction



(P-Cl$_5$+H) of $^{13}$C$_6$-HCB possesses the identical chemical formula as that of the molecular ion of $^{13}$C$_6$-pentachlorobenzene, and thus they cannot be differentiated by HRMS even though isotopologue distribution comparison is employed (Table S8). Likewise, the ion P-Br$_5$+H of $^{13}$C$_6$-HBB cannot be distinguished from the molecular ion of $^{13}$C$_6$-pentabromobenzene by HRMS in combination with isotopologue distribution comparison. As a result, the dechlorination/debromination ions with H-abstraction of the HOCs containing more chlorine/bromine atoms could trigger interference to analysis of those containing less chlorine/bromine atoms when HRMS and isotopologue distribution comparison are employed, possibly leading to false positive identification results. Since the H-abstraction efficiencies of some dechlorination/debromination ions are relatively high (e.g., the P-Br$_5$ of $^{13}$C$_6$-HBB), the interference may be considerable. Therefore, other separation techniques such as GC and ion-mobility MS are necessary to couple with HRMS for identification of HOCs, thus eliminating the interferences and improving the validity of analysis results. However, in some scenarios, the GC retention time ranges of HOC congeners containing different Cl/Br atoms may partially overlap. Therefore, more data such as chlorine/bromine isotopologue distribution patterns and quantitative structure-retention relationships are warranted for getting rid of the possible interference caused by H-abstraction reactions on EI-MS.[60] In addition, when using EI-MS data to evaluate the relative abundances of carbon isotopologues and carbon isotope ratios of HOCs, analysts should take into account the contribution of signal from the ions with H-abstraction to the carbon isotopologues containing one $^{13}$C atom.



# 4 CONCLUSIONS

In this work, we systematically investigated the H-abstraction reactions of $^{13}C_6$-HCB and $^{13}C_6$-HBB on EI source using GC-HRMS. The H-abstraction efficiencies were assessed with the MS signal intensity ratios of the ions with H-abstraction relative to the corresponding original ions. Three levels of H-abstraction efficiencies, that is, H-abstraction efficiency of each isotopologue, whole H-abstraction efficiency of each ion, and overall H-abstraction efficiency of each compound were calculated. For most ions, the H-abstraction efficiencies of individual isotopologues declined from the first to the last isotopologues. The whole H-abstraction efficiencies of individual ions can be distinguished from each other for individual compounds, and the overall H-abstraction efficiency of $^{13}C_6$-HCB was significant lower than that of $^{13}C_6$-HBB. In addition, the isotopologue distributions of measured MS signal intensities between $[M]^{\bullet+}$ and $[M+H]^+$ for each compound were evidently different, and those of the latter contradicted the distributions of theoretical relative abundances of isotopologues. The impacts of several instrumental parameters and factors including temperatures of ion source, EI energies and numbers of $^{37}Cl$ or $^{81}Br$ atoms of isotopologues on the H-abstraction efficiencies were revealed. The increase of ion source temperatures firstly enhanced the H-abstraction efficiencies to a maximum and then began to linearly reduce the efficiencies. The H-abstraction efficiencies were significantly affected by the EI energies and emission currents, and reached the highest at the EI energy of 45 eV and emission current of 0.9 mA. The mechanisms underlying the H-abstraction reactions of HOCs on EI source are tentatively proposed in light of the previous studies and evidences found in the current study. The H-abstraction reactions



might pertain to ion-molecule reactions, and the siloxanes running off from the GC column possibly participated in the reactions as a hydrogen source. Furthermore, the H-abstraction reactions might be kinetically controlled, and chlorine and bromine isotopic effects were observed in the reactions. This study sheds light on the potential mechanisms and magnitudes of the H-abstraction reactions of HOCs on EI source, and will facilitate the qualitative analysis of HOPs in environmental research, particularly for identification of novel HOPs. In the future, it will be worthwhile to apply computational quantum chemistry to unveiling the in-depth mechanisms behind the EI-induced H-abstraction reactions of HOCs. The details of the possible hydrogen donor, siloxanes bleeding from the GC column, also merit further investigation.



# SUPPORTING INFORMATION

The Supporting Information is available on the website at http://pending.

# ACKNOWLEDGEMENTS

This study was financially supported by the National Natural Science Foundation of China (Grant No. 41603092).

# CONFLICT OF INTEREST

The authors have no conflict of interest to declare.

**Figure legends**

**Figure 1**. Representative chromatograms and high resolution mass spectra of $^{13}C_6$-hexachlorobenzene ($^{13}C_6$-HCB) and $^{13}C_6$-hexabromobenzene ($^{13}C_6$-HBB). P-Cl$_n$: dechlorination product ion possessing n Cl atom(s); P-Br$_n$: debromination product ion possessing n Br atom(s); NL: nominal level; *m/z*: mass to charge ratio.

**Figure 2**. Hydrogen-abstraction efficiencies ([M+H]$^+$/[M]$^{•+}$ or [P+H]$^+$/[P]$^+$ MS signal intensity ratios) of chlorine/bromine isotopologues of molecular and dechlorination/debromination product ions of $^{13}C_6$-HCB and $^{13}C_6$-HBB on electron ionization high-resolution mass spectrometry (EI-HRMS). [M]$^{•+}$: a molecular ion; [M+H]$^+$: the corresponding molecular ion with hydrogen abstraction; [P]$^+$: a dehalogenation product ion; [P+H]$^+$: the corresponding product ion with H-abstraction; IST-n: an isotopologue of an ion possessing n−1 $^{37}Cl$ or $^{81}Br$ atom(s). Error bars denote the standard deviations (1 σ, n = 6).

**Figure 3**. Whole hydrogen-abstraction efficiencies of individual ions and overall hydrogen-abstraction efficiencies of $^{13}C_6$-HCB and $^{13}C_6$-HBB on EI-HRMS.

**Figure 4**. Distribution patterns of MS signal intensities of chlorine/bromine isotopologues of the molecular ions ([M]$^{•+}$) and the molecular ions with hydrogen abstraction ([M+H]$^+$) of $^{13}C_6$-HCB and $^{13}C_6$-HBB on GC-EI-HRMS.

**Figure 5**. Plots of the whole H-abstraction efficiencies of the molecular ion ([M]$^{•+}$) of $^{13}C_6$-HCB versus **(a)** the ion source temperatures, **(b)** the reciprocal Kelvin temperatures of ion



source, **(c)** the EI energies, and **(d)** the emission currents of EI filament, as well as the correlations of **(e)** the H-abstraction efficiencies versus the varied MS signal intensities caused by different ion source temperatures and **(f)** the H-abstraction efficiencies versus the varied MS signal intensities trigged by different EI energies. The injection replicates was three.

**Figure 6**. Correlations between H-abstraction efficiencies and the $^{37}$Cl or $^{81}$Br atom numbers of chlorine/bromine isotopologues. **(a)**: $[M+H]^+/[M]^{\bullet+}$ intensity ratios of the chlorine isotopologues of the molecular ion of $^{13}C_6$-HCB; **(b)**: $[P+H]^+/[P]^+$ intensity ratios of the chlorine isotopologues of the P-Cl$_5$ of $^{13}C_6$-HCB; **(c)**: $[M+H]^+/[M]^{\bullet+}$ intensity ratios of the bromine isotopologues of the molecular ion of $^{13}C_6$-HBB; **(d)**: $[P+H]^+/[P]^+$ intensity ratios of the bromine isotopologues of the P-Br$_5$ of $^{13}C_6$-HBB; solid curves refer to exponential regressions and shaded areas represent the corresponding 95% confidence intervals.



# Figures

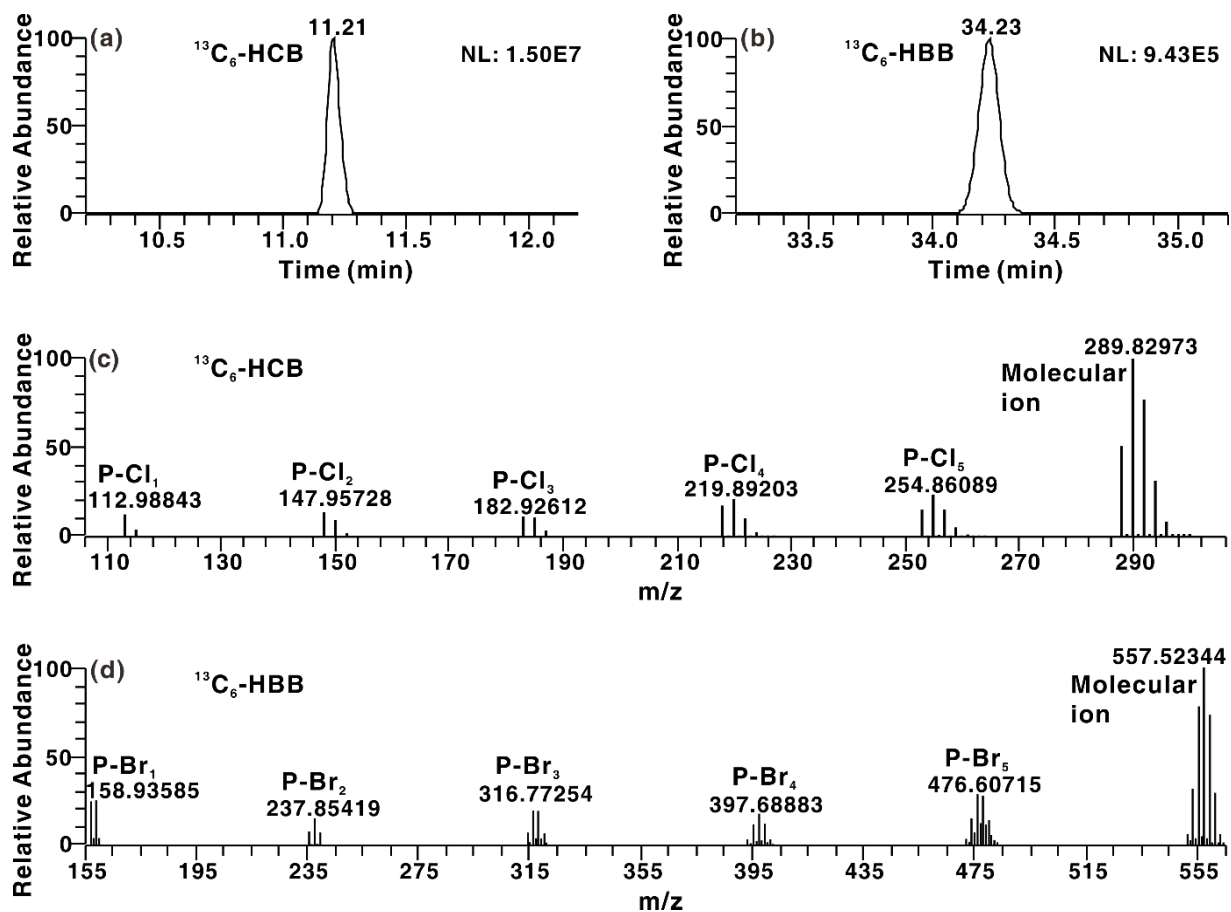

**Figure 1**. Representative chromatograms and high resolution mass spectra of $^{13}C_6$-hexachlorobenzene ($^{13}C_6$-HCB) and $^{13}C_6$-hexabromobenzene ($^{13}C_6$-HBB). P-Cl$_n$: dechlorination product ion possessing n Cl atom(s); P-Br$_n$: debromination product ion possessing n Br atom(s); NL: nominal level; $m/z$: mass to charge ratio.



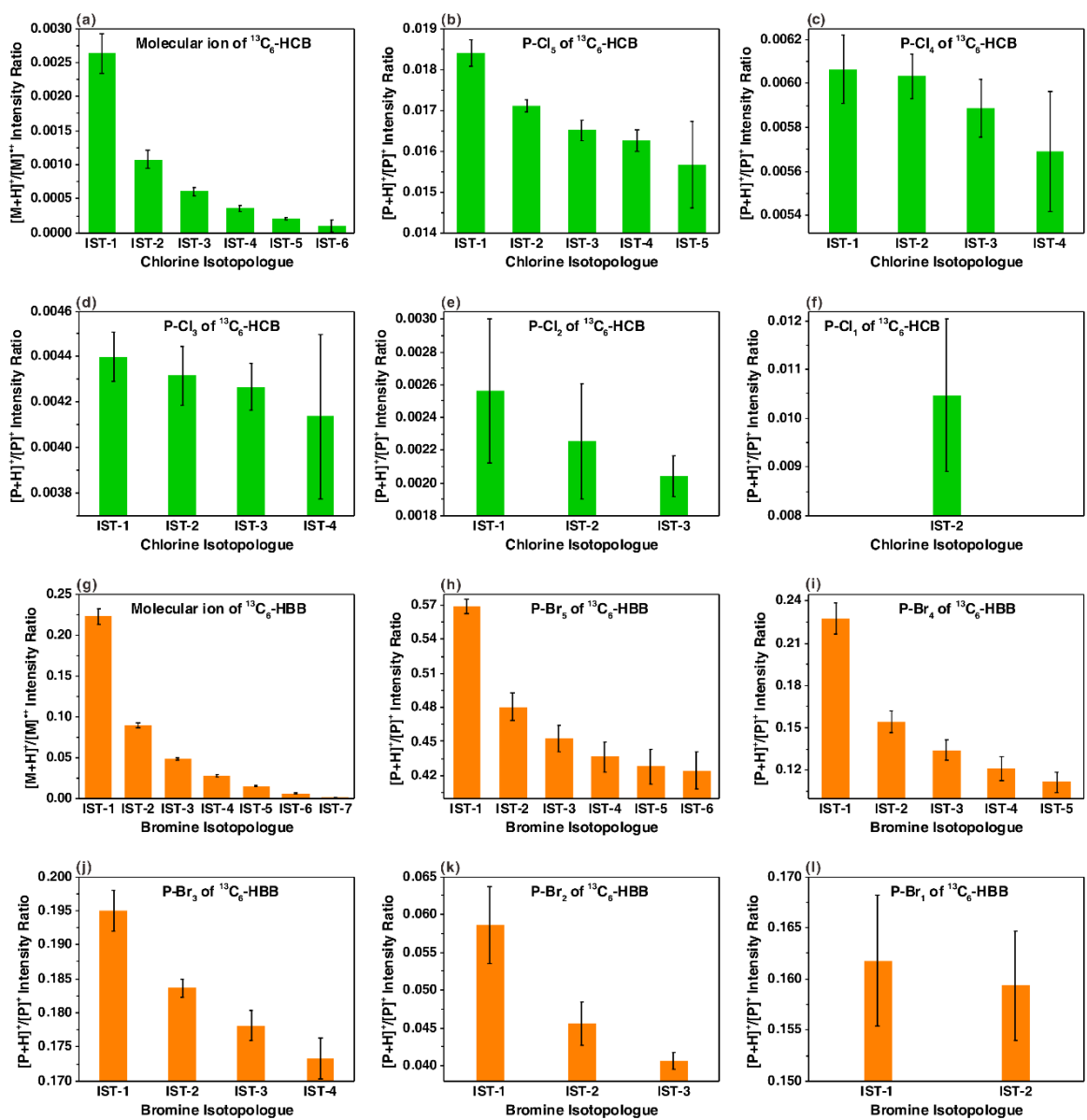

**Figure 2**. Hydrogen-abstraction efficiencies ([M+H]$^+$/[M]$^{+\bullet}$ or [P+H]$^+$/[P]$^+$ MS signal intensity ratios) of chlorine/bromine isotopologues of molecular and dechlorination/debromination product ions of $^{13}C_6$-HCB and $^{13}C_6$-HBB on electron ionization high-resolution mass spectrometry (EI-HRMS). [M]$^{+\bullet}$: a molecular ion; [M+H]$^+$: the corresponding molecular ion with hydrogen abstraction; [P]$^+$: a dehalogenation product ion; [P+H]$^+$: the corresponding product ion with H-abstraction; IST-n: an isotopologue of an ion possessing n−1 $^{37}$Cl or $^{81}$Br atom(s). Error bars denote the standard deviations (1 σ, n = 6).



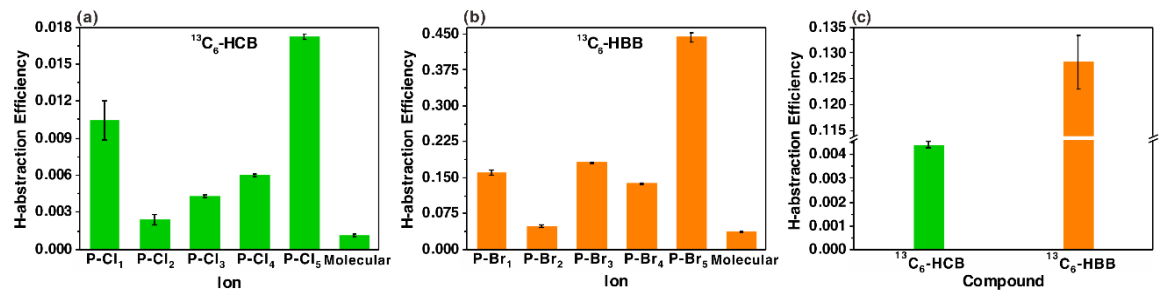

**Figure 3**. Whole hydrogen-abstraction efficiencies of individual ions and overall hydrogen-abstraction efficiencies of $^{13}C_6$-HCB and $^{13}C_6$-HBB on EI-HRMS.



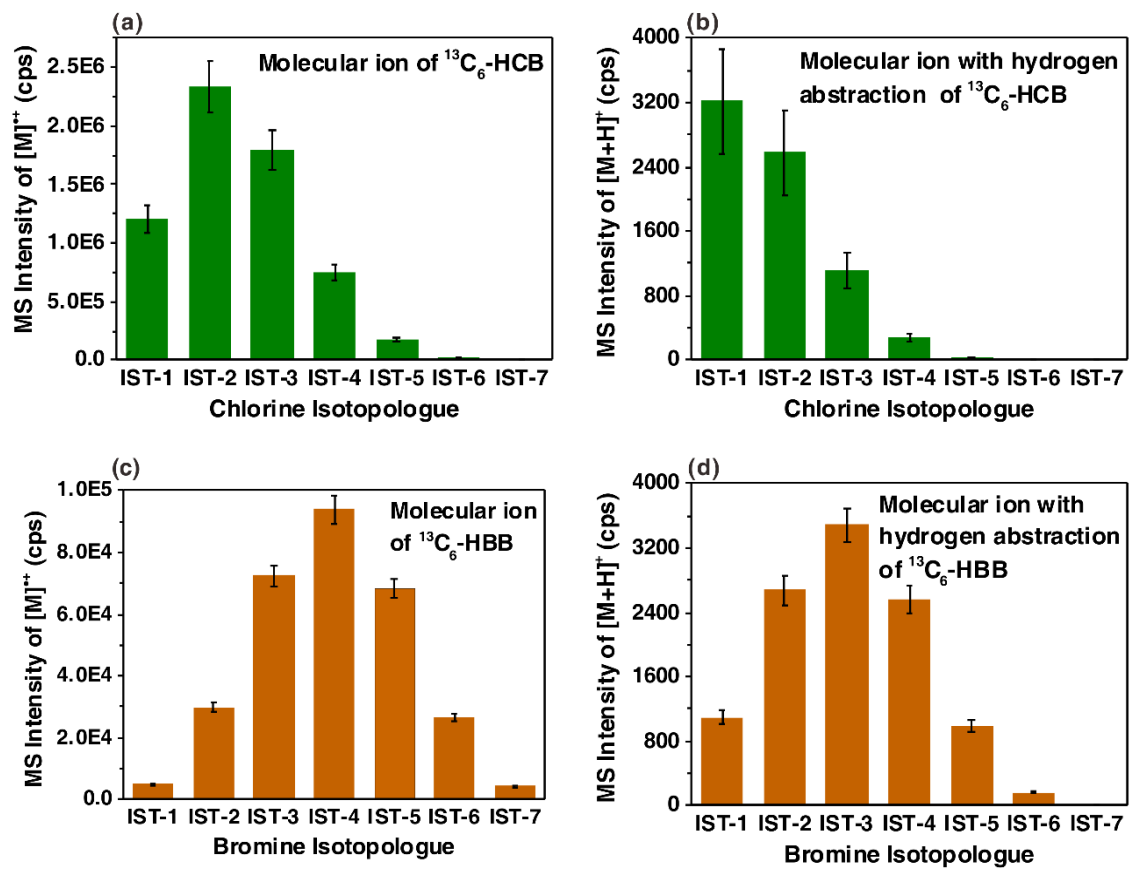

**Figure 4**. Distribution patterns of MS signal intensities of chlorine/bromine isotopologues of the molecular ions ([M]$^{\bullet+}$) and the molecular ions with hydrogen abstraction ([M+H]$^{+}$) of $^{13}C_6$-HCB and $^{13}C_6$-HBB on GC-EI-HRMS.



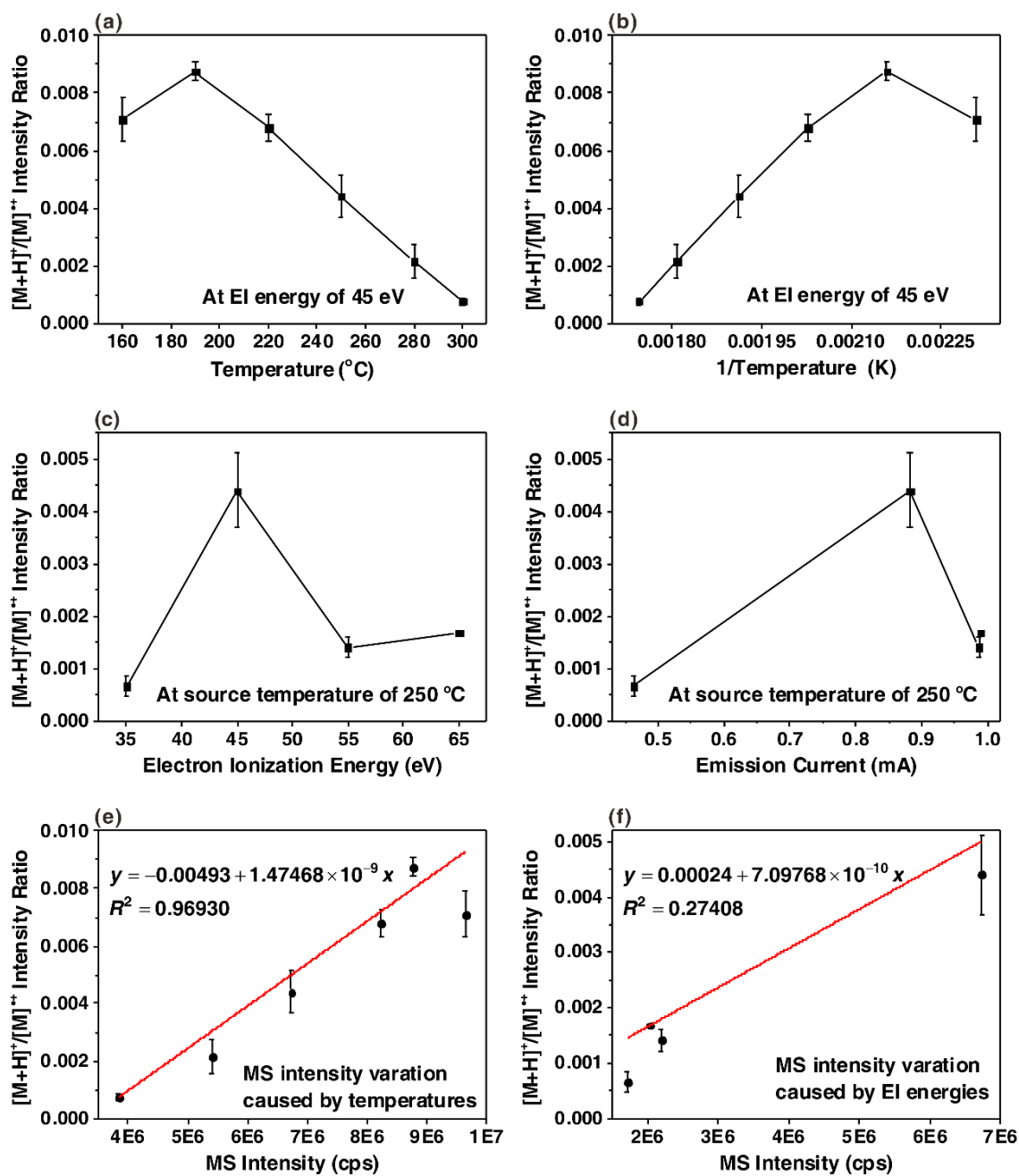

**Figure 5**. Plots of the whole H-abstraction efficiencies of the molecular ion ($[M]^{+\bullet}$) of $^{13}C_6$-HCB versus (**a**) the ion source temperatures, (**b**) the reciprocal Kelvin temperatures of ion source, (**c**) the EI energies, and (**d**) the emission currents of EI filament, as well as the correlations of (**e**) the H-abstraction efficiencies versus the varied MS signal intensities caused by different ion source temperatures and (**f**) the H-abstraction efficiencies versus the varied MS signal intensities triggered by different EI energies. The injection replicates was three.



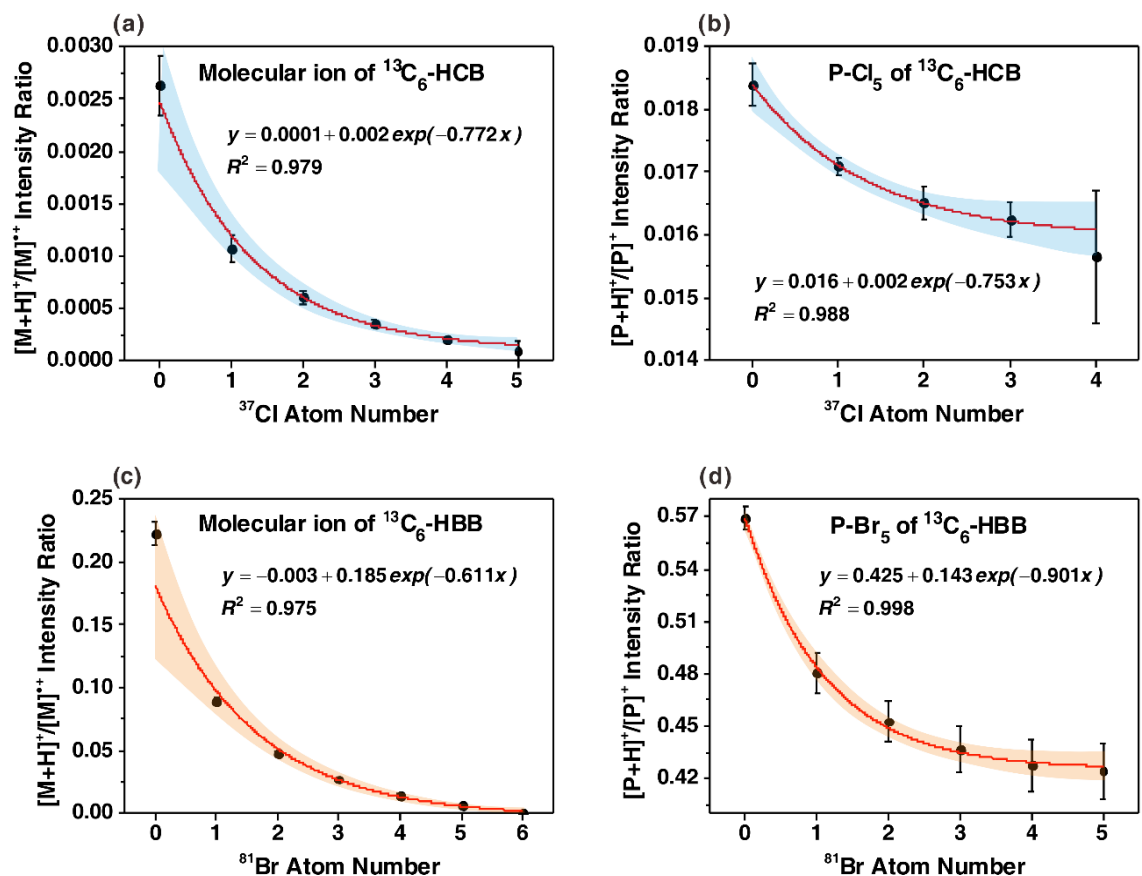

**Figure 6**. Correlations between H-abstraction efficiencies and the $^{37}$Cl or $^{81}$Br atom numbers of chlorine/bromine isotopologues. **(a)**: [M+H]$^+$/[M]$^{•+}$ intensity ratios of the chlorine isotopologues of the molecular ion of $^{13}$C$_6$-HCB; **(b)**: [P+H]$^+$/[P]$^+$ intensity ratios of the chlorine isotopologues of the P-Cl$_5$ of $^{13}$C$_6$-HCB; **(c)**: [M+H]$^+$/[M]$^{•+}$ intensity ratios of the bromine isotopologues of the molecular ion of $^{13}$C$_6$-HBB; **(d)**: [P+H]$^+$/[P]$^+$ intensity ratios of the bromine isotopologues of the P-Br$_5$ of $^{13}$C$_6$-HBB; solid curves refer to exponential regressions and shaded areas represent the corresponding 95% confidence intervals.